# CHAOS STRUCTURES. MULTICURRENCY ADVISER ON THE BASIS OF NSW MODEL AND SOCIAL-FINANCIAL NETS

A.M.Avdeenko

National University of Science and Technology «Moscow Institute of Steel and Alloys», Moscow, Russia

119049, Moscow, Leninsky prospekt, 4

e-mail: aleksei-avdeenko@mail.ru

## 1. Introduction

Certainly, professionals have some reasons to muzz customers of stock strategies. The more complicated system brings more difficult formalization of description space and laws of system evolution. At the same time successful choice of description space allows finding laws, which control system behavior, with minimum costs.

In the present work it is proposed an algorithm of multicurrency strategy optimization for Forex market, however it can be used both for the other markets of shares, options, futures etc. Nonlinear stochastic wavelet model (NSW model), stated in the works [1, 2] has been used as a basis of the algorithm (so called elementary decision units).

Within the framework of this model, the stochastic process $X(t,\omega)$ is the price of shares, options and futures; and indication of price of currency couples at the Forex market within the time point $t \subset R$, where $\omega \subset \Omega$ is a point of probability space with measure function $P$, appears as population of analyzing function-wavelets $\psi_{i\tau}(t) = 2^{-i/2}\psi(2^{-i}t - \tau)$, which form orthonormal basis for $L^2(R)$ with compact carrier.

Stochastic function $X(t,\omega)$ is represented in the form of linear combination of functions $Y_i(\tau) = \int X(t,\omega)\psi_{i\tau}(t)dt$ and it is in accord with Ito stochastic equation $d\mathbf{Y}(\tau) = \mathbf{F}(\mathbf{Y}(\tau))d\tau + \mathbf{G}(\mathbf{Y}(\tau))d\boldsymbol{\omega}$, where $d\omega_i(t)$ is the random Wiener process $<d\omega_i(t)> = 0$, $<d\omega_i(t)d\omega_j(t_1)> = \delta_{ij}\delta(t-t_1)$, $i,j = 1...J$, and $\mathbf{F}(\mathbf{Y}(\tau)), \mathbf{G}(\mathbf{Y}(\tau))$ are drift and diffusional components respectively; and we set $G_{ij}(\mathbf{Y}(\tau)) = \delta_{ij}G(\mathbf{Y}(\tau))$.

Naturally, the model has non-Markov nature. However, in single-mode case $J=1$ time independent probability density, find as $P(Y_1 < y_1) = \int_{-\infty}^{y_1} f_s(y_1)dy_1$, can be represented in the form of quadratures: $f_s(y_1) \approx e^{W(y_1)}$, where $W(y_1) = \int_0^{y_1} \frac{2F_1(y_1)}{G_1^2(y_1)} dy_1$.

Henceforth, the motion equation and corresponding probabilities are restored according to inquiries by means of expansion in a series by suitable set of orthogonal functions (in our case they are Ermit polynoms) using extremum principle.

So, in elementary case, the moment of optimal purchase (entry into the long position) can be represented in the form $t = \inf(-dy_1 > 0, P_s > 1 - \alpha_1)$ and sales (entry into the short position) $t = \inf(-dy_1 < 0, P_s < \alpha_1)$, where $\alpha_1$ is risk level, $P_s = \int_{-\infty}^{0} f_s(y)dy$.

Alternative choice is to use the convolution $f_s(z) = \int_{-\infty}^{+\infty} f_s(y_1) f_{s,T}(y_1 + z)dy_1$. Here the conditions $P_s > 1 - \alpha_1, P_s < \alpha_1$ are the intent of resale and repurchase moments.

In both cases it is necessary to exclude statistically nonsignificant differences between stationary distributions in time shifts $t \to t+T$, and it is fulfilled using Kolmogorov-Smirnov nonparametric criterion.

Henceforth we shall name criteria associated with the sign of $dy_1$ as elementary dynamic ones and ones with probabilistic assessments as elementary statistical criteria.

## 2. "Chaos Structures" Model

The algorithm proposed is suspension over elementary models, which permit as far as it possible to simulate effective decision-making. The control flow chart is given in Figure 1. Elementary decision generators are specified as 1 and 2, and they conform to statistical and dynamic work criteria [2].

Total unit number is not limited; and here standard elementary models, such as Moving Average Convergence/Divergence (MACD) method, Bollinger Bands (BB), Market Relative Strength Index etc., can be used.

The following principles are used as a basis of the algorithm:

1. Dynamic optimization with the possibility of generation of random solutions, which "survive" or "do not survive" depending on their effectiveness ("creativeness"), unit 4. Here Boolean algebra elements are used, i.e. weighting coefficient of elementary decisions for entry/ exit in to short or long position is taken in such a manner as to minimize norm function of differences between sign of quotation change and decision made as per dynamic and statistical criteria.

The necessity of generation of random solutions nearby local optimums is associated with nonidentity of local efficiency for given currency couple (e.g. maximum profitability of trading within the specified period) to global optimality/ profitability for the package of all currency couples.

2. Possibility of tightly- and loosely-coupled horizontal self-assemblies, i.e. automatic connection or disconnection of various program units, directed on various currency couple trading (Unit 6).

The decision, made in Unit 4 for every currency couple, is summarized considering coupled correlation functions. For example, if two currency couples have positive correlation coefficient, then the decision made in Unit 4 regarding entry in to short position as per one currency couple increase probability similar to decision per the other currency couple and vice versa in negative correlation.

At that the units can be in different states that are active trading (decision-making and mutual interaction with trade server - Unit S), semi-active state (discussion of decisions and interchange of information with the other units) and passive state (when the units do not take part in decision making and discussion). Transition between the states is determined for each currency couple as per the quality of prediction of the future quotation change (Unit 7).

Tightly-coupled self-assembly enables to make the collective decisions concerning market entry/ exit on the basis of the current quotations in specified time scale, and in case of loosely-coupled self-assembly - on the basis of effectiveness of completed activity in various time scales, Unit 8.

Loosely-coupled self-assembly simulates realization of mixed strategy of work [2], i.e. compromise between maximum profitability and minimum risk: package reallocation takes place (parts of means invested in short or long positions of various currency couples), Unit 5.

3. Possibility of self-homothetic assembly enables to use the whole structure completely as the unit of elementary decision making, but in the other time scale (with other time frame), Unit 3.

4. Possibility of vertical self-assembly: at a time the decision achieved (specifically, entry or exit in short or long position) can be used as the entry variables along with current quotations, Unit N. Some times it allows make strategic decisions, however performance requirements to computing machinery grow up sharply.

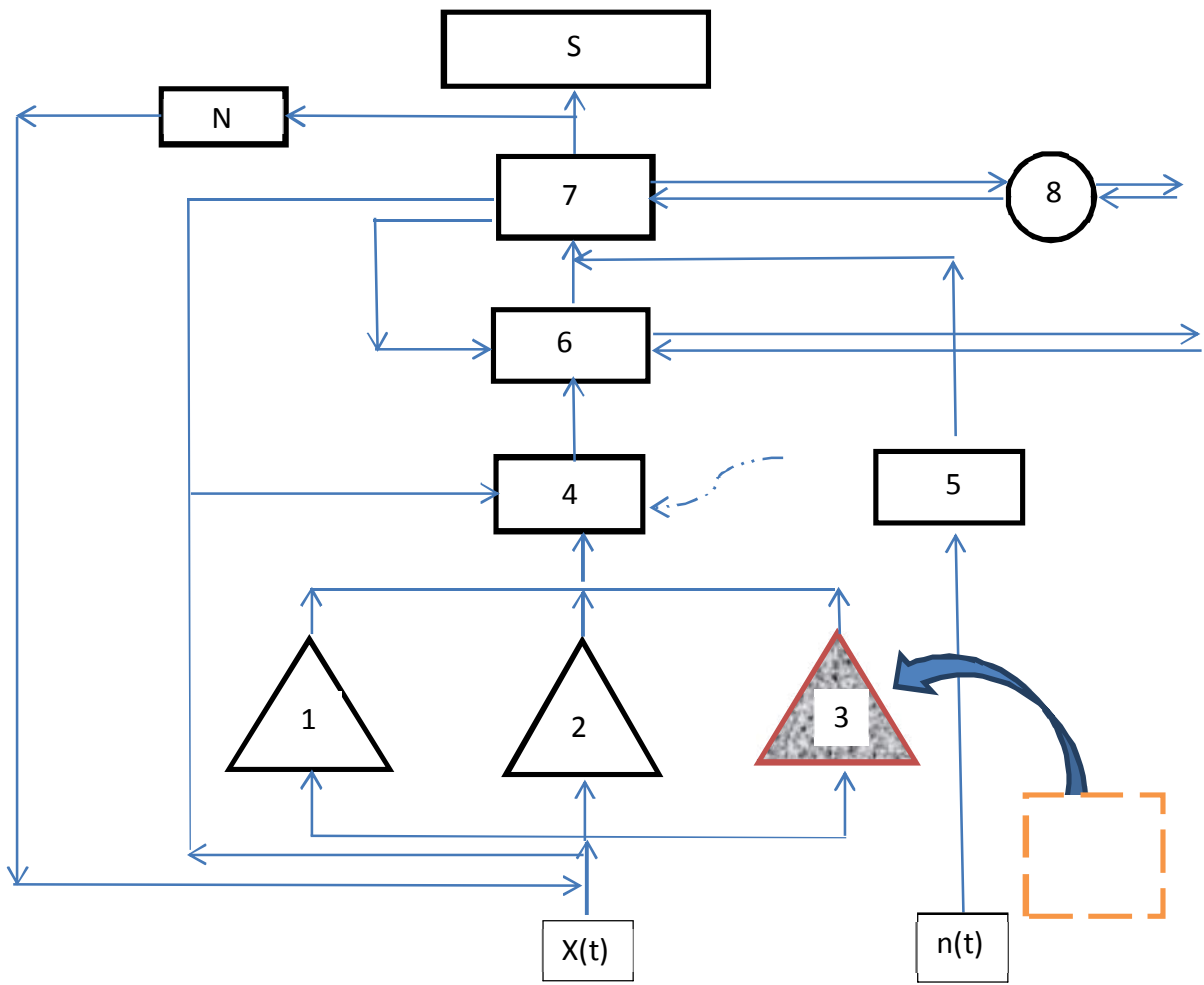

Fig.1 Basic diagram of the "Chaos Structures" algorithm of NSW model.

1, 2.- elementary decision generators (NSW model); 3-unit of self-homothetic assembly; 4- unit of dynamic optimization and structure generation; 5- package optimizer; 6-collective decision unit; 7- unit of tightly-coupled self-assembly generation; 8- unit of loosely-coupled self-assembly generation; N-unit of vertical self-assembly; S- central trade server; X(t), n(t)- current quotation and part of currency couple in financial instrument package.

## 3. Experimental validation

Experimental validation was carried out in the mode of real-time automatic vending within the period April 28, 2011 to May 26, 2011, when simultaneous trading with eight currency couples with timeframe of 5 minutes using single self-homothetic assembly with timeframe of 15 minutes. The algorithm was realized in MetaTrader medium. In elementary variant its value amounted about one and a half thousand program lines.

Vertical self-assembly was not used. Initial deposit amounted $5000. Detailed results are represented in Table 1 and in Supplement 1. There was not any preliminary adjustment and fitting of model parameters and the algorithm.

Table 1. Operating effectiveness of "Chaos Structures" algorithm within the period of April 28, 2011 to May 26, 2011 (In the brackets there is the number of lucrative transactions.)

| Deposit $; | Number of transactions | | Closed P/L,$ | Floating P/L,$ | Average profit per transaction $; |
|---|---|---|---|---|---|
| | Long | Short | | | |
| 5000 | 83 (83) | 122(122) | 5683.62 | -1 161.37 | 27.87 |

## 4. Discussion

According to the author the proposed algorithm shows high efficiency due to algorithm flexibility. It is not possible to predict, in particular, what units will be active and how they will be active, as well as considering what time scale the decisions will be made. Elementary upper estimate shows that for M units of elementary decisions and N currency couples under K-multiple self-homothetic investment total number of possible communications between various currency couples in various timeframes have the order of $[(M!)(N!)]^{K+1}$.

For the above mentioned M=2, N=8, and K=1 this value amounts $6*10^9$ as per the order. Inclusion of vertical self-assembly increases this estimate yet more.

The proposed algorithm is not one or another variant of neural net model. The neural nets has shown absolute ineffectiveness in solution of these tasks, because currency couple quotation forming is essentially random and possible with application of determinate chaos effect in one or another model, from Feigenbaum, to Lorentz with probable inclusion of Y. Pomeau and P. Manneville scenarios.

Volatility of the process is irreducible to one time-scale, conversely there are multi-scale phenomena; so net "training" is useless or leads to trivial results.

Algorithms of elementary solution unit [2] are rather close to stochastic approximation procedures of Robbins-Monro type [3]. The system, as a whole, rather simulates the "human" method of decision making: from accumulation and systematization of available information to development of various level models followed by clarification and possibility of "afflatus". One, but essential, difference is much more scope of processing information and speed.

## 5. Forex market super profit and social-and-financial nets

Serious disadvantage of this model of programme trading at the Forex market or the other markets is the circumstance that the adviser can only follow the market cost fluctuations changing the part of one or another currency couple in the package due to loosely-coupled self-assembly (mixed strategy). At the same time there is the method of cardinal change of effectiveness of on-line trading on the basis of stochastic synchronization ideas.

Fantastic success of social nets such as Facebook, Twitter and others gives us this idea. Indeed, simultaneous (usually within the limits of the minute timeframe) entrance in to the market in short or long position of about 500-1000 standard lots (the average cost of each is $100 000) or 2-3 thousand million dollars per 23 negotiated currency couples leads to essential modifications of currency couple quotations.

This is the upper estimate. Even if the lever is 10, such sums are unreachable for 95 percent of small and medium investors in the time of individual commercial business.

However, simple estimates show that if only it was possible within the limits of the minute timeframe to synchronize trade flows within the framework of abovementioned algorithm for several dozens of thousand small investors, then not great investors, but just they would determine to a great extent the market quotations. The system of self-fulfilling predictions would arise. The market would become more stable and predictable and it would reduce the probability of financial accidents and sovereign defaults.

Project implementation is possible via development of social-financial net, entry to which attaches user terminal to central server synchronizing the decision-making on the basis of proposed algorithm. Further development of this idea means involvement in a like manner of dozens of millions of extra-small investors via mobile network. Great cellular operators can be interested in such diversification of services. Most probably in the nearest future these projects will be proposed and realized.

6. Reference literature

Annotation

Algorithm of multicurrency trading at the market of Forex is realized on the basis of nonlinear stochastic wavelets. The distinctive feature of the algorithm is the possibility of weakly- and strongly connected horizontal self-assemblies, as well as use of nested structures. On-line trading with eight currency couples has shown high effectiveness and stability of the algorithm.

It is discussed the problem of possibility of excess profit earning in electronic markets via development of social-financial nets based on synchronization of work of individual traders by means of proposed algorithm.

Summary

Algorithm of multicurrency trading at the market of Forex is realized on the basis of nonlinear stochastic wavelets. The following principles are used as a basis of the algorithm:

1. Dynamic optimization with the possibility of generation of random solutions, which "survive" or "do not survive" depending on their effectiveness.

2. Possibility of tightly- and loosely-coupled horizontal self-assemblies, i.e. automatic connection or disconnection of various program units, directed on various currency couple trading. Tightly-coupled self-assembly enables to make the collective decisions concerning market entry/ exit on the basis of the current quotations in specified time scale, and in case of loosely-coupled self-assembly - on the basis of effectiveness of completed activity in various time scales.

3. Possibility of self-homothetic assembly enables to use the whole structure completely as the unit of elementary decision making, but in the other time scale.

On-line trading with eight currency couples has shown high effectiveness and stability of the algorithm.

It is discussed the problem of possibility of excess profit earning in electronic markets via development of social-financial nets based on synchronization of work of individual traders by means of proposed algorithm.

Suppl.1

Closed Transactions:

| Ticket | Open Time | Type | Size | Item | Price | S / L | T / P | Close Time | Price | Commission | Taxes | Swap | Profit |
|---|---|---|---|---|---|---|---|---|---|---|---|---|---|
| 111018750 | 2011.04.01 11:19 | balance | | Deposit | | | | | | | | | 5000.00 |
| 114266871 | 2011.04.28 09:14 | sell | 1.00 | gbpusd | 1.66839 | 0.00000 | 0.00000 | 2011.04.28 09:23 | 1.66819 | 0.00 | 0.00 | 0.00 | 20.00 |
| 114266891 | 2011.04.28 09:14 | buy | 0.17 | usdchf | 0.87083 | 0.78070 | 0.87520 | 2011.04.28 09:44 | 0.87115 | 0.00 | 0.00 | 0.00 | 6.24 |
| 114266901 | 2011.04.28 09:14 | buy | 0.16 | audusd | 1.09161 | 1.09226 | 1.09589 | 2011.04.28 09:21 | 1.09252 | 0.00 | 0.00 | 0.00 | 14.56 |
| 114266908 | 2011.04.28 09:14 | buy | 0.15 | eurusd | 1.48428 | 1.48455 | 1.48868 | 2011.04.28 09:18 | 1.48455 | 0.00 | 0.00 | 0.00 | 4.05 |
| 114267720 | 2011.04.28 09:18 | sell | 0.12 | eurusd | 1.48450 | 1.48361 | 1.48015 | 2011.04.28 09:51 | 1.48361 | 0.00 | 0.00 | 0.00 | 10.68 |
| 114268688 | 2011.04.28 09:21 | sell | 0.15 | audusd | 1.09253 | 1.18270 | 1.08820 | 2011.04.28 19:09 | 1.08820 | 0.00 | 0.00 | 0.00 | 64.95 |
| 114268690 | 2011.04.28 09:21 | buy | 0.15 | audusd | 1.09270 | 1.09282 | 1.09703 | 2011.04.28 09:29 | 1.09311 | 0.00 | 0.00 | 0.00 | 6.15 |
| 114269023 | 2011.04.28 09:23 | sell | 0.23 | gbpusd | 1.66803 | 1.66757 | 1.66373 | 2011.04.28 09:52 | 1.66757 | 0.00 | 0.00 | 0.00 | 10.58 |
| 114273055 | 2011.04.28 09:44 | sell | 0.22 | usdchf | 0.87120 | 0.96134 | 0.86684 | 2011.04.29 12:19 | 0.86684 | 0.00 | 0.00 | -0.13 | 110.65 |
| 114273057 | 2011.04.28 09:44 | buy | 0.22 | usdchf | 0.87134 | 0.78120 | 0.87570 | 2011.04.28 14:06 | 0.87570 | 0.00 | 0.00 | 0.00 | 109.54 |
| 114274647 | 2011.04.28 09:51 | buy | 0.22 | eurusd | 1.48362 | 1.48378 | 1.48802 | 2011.05.02 12:29 | 1.48378 | 0.00 | 0.00 | 0.62 | 3.52 |
| 114274908 | 2011.04.28 09:52 | sell | 0.22 | gbpusd | 1.66746 | 1.66724 | 1.66316 | 2011.04.28 09:55 | 1.66724 | 0.00 | 0.00 | 0.00 | 4.84 |
| 114275554 | 2011.04.28 09:55 | sell | 0.22 | gbpusd | 1.66705 | 1.75725 | 1.66275 | 2011.04.29 03:09 | 1.66275 | 0.00 | 0.00 | -0.77 | 94.60 |
| 114578899 | 2011.05.02 11:46 | sell | 0.26 | gbpusd | 1.66873 | 1.66845 | 1.66440 | 2011.05.02 11:49 | 1.66815 | 0.00 | 0.00 | 0.00 | 15.08 |
| 114579239 | 2011.05.02 11:49 | sell | 0.23 | usdchf | 0.86877 | 0.95892 | 0.86442 | 2011.05.02 12:13 | 0.86859 | 0.00 | 0.00 | 0.00 | 4.77 |
| 114579281 | 2011.05.02 11:49 | sell | 0.22 | audusd | 1.09512 | 1.09505 | 1.09080 | 2011.05.02 13:18 | 1.09481 | 0.00 | 0.00 | 0.00 | 6.82 |
| 114579314 | 2011.05.02 11:49 | sell | 0.23 | gbpusd | 1.66789 | 1.66776 | 1.66358 | 2011.05.02 11:56 | 1.66776 | 0.00 | 0.00 | 0.00 | 2.99 |
| 114579993 | 2011.05.02 11:56 | sell | 0.23 | gbpusd | 1.66757 | 1.66727 | 1.66328 | 2011.05.02 12:00 | 1.66727 | 0.00 | 0.00 | 0.00 | 6.90 |
| 114580584 | 2011.05.02 12:00 | sell | 0.23 | gbpusd | 1.66712 | 1.66702 | 1.66278 | 2011.05.02 12:04 | 1.66702 | 0.00 | 0.00 | 0.00 | 2.30 |
| 114580896 | 2011.05.02 12:04 | sell | 0.23 | gbpusd | 1.66695 | 1.66682 | 1.66262 | 2011.05.02 12:09 | 1.66652 | 0.00 | 0.00 | 0.00 | 9.89 |

| | | | | | | | | | | | | |
|---|---|---|---|---|---|---|---|---|---|---|---|---|
| 114581300 | 2011.05.02 12:09 | sell | 0.23 | gbpusd | 1.66634 | 1.75649 | 1.66199 | 2011.05.03 03:42 | 1.66199 | 0.00 | 0.00 | -0.87 | 100.05 |
| 114581619 | 2011.05.02 12:13 | sell | 0.23 | usdchf | 0.86832 | 0.95859 | 0.86409 | 2011.05.02 12:26 | 0.86808 | 0.00 | 0.00 | 0.00 | 6.36 |
| 114583156 | 2011.05.02 12:26 | sell | 0.23 | usdchf | 0.86791 | 0.86769 | 0.86360 | 2011.05.02 12:31 | 0.86717 | 0.00 | 0.00 | 0.00 | 19.63 |
| 114583655 | 2011.05.02 12:29 | sell | 0.23 | eurusd | 1.48375 | 1.48367 | 1.47940 | 2011.05.02 13:11 | 1.48340 | 0.00 | 0.00 | 0.00 | 8.05 |
| 114583659 | 2011.05.02 12:29 | sell | 0.23 | eurusd | 1.48375 | 1.48369 | 1.47940 | 2011.05.02 13:11 | 1.48343 | 0.00 | 0.00 | 0.00 | 7.36 |
| 114584423 | 2011.05.02 12:31 | sell | 0.20 | usdchf | 0.86695 | 0.95717 | 0.86267 | 2011.05.02 14:14 | 0.86614 | 0.00 | 0.00 | 0.00 | 18.70 |
| 114589185 | 2011.05.02 13:11 | sell | 0.24 | eurusd | 1.48331 | 1.57341 | 1.47891 | 2011.05.03 02:47 | 1.47891 | 0.00 | 0.00 | -1.80 | 105.60 |
| 114589759 | 2011.05.02 13:18 | sell | 0.23 | audusd | 1.09463 | 1.18480 | 1.09030 | 2011.05.03 05:03 | 1.09030 | 0.00 | 0.00 | -4.55 | 99.59 |
| 114594534 | 2011.05.02 14:14 | sell | 0.23 | usdchf | 0.86606 | 0.86591 | 0.86170 | 2011.05.02 14:31 | 0.86591 | 0.00 | 0.00 | 0.00 | 3.98 |
| 114596770 | 2011.05.02 14:31 | sell | 0.22 | usdchf | 0.86572 | 0.95592 | 0.86142 | 2011.05.03 17:55 | 0.86142 | 0.00 | 0.00 | -0.13 | 109.82 |
| 115372685 | 2011.05.05 20:32 | sell | 1.15 | eurusd | 1.45699 | 1.45623 | 1.45259 | 2011.05.05 20:37 | 1.45623 | 0.00 | 0.00 | 0.00 | 87.40 |
| 115372738 | 2011.05.05 20:32 | sell | 0.22 | usdchf | 0.86963 | 0.95975 | 0.86525 | 2011.05.05 20:58 | 0.86940 | 0.00 | 0.00 | 0.00 | 5.82 |
| 115372779 | 2011.05.05 20:33 | buy | 0.21 | audusd | 1.06376 | 0.97359 | 1.06809 | 2011.05.06 04:41 | 1.06809 | 0.00 | 0.00 | 1.66 | 90.93 |
| 115373366 | 2011.05.05 20:36 | buy | 0.20 | gbpusd | 1.64186 | 1.64221 | 1.64603 | 2011.05.06 13:54 | 1.64221 | 0.00 | 0.00 | -0.08 | 7.00 |
| 115373514 | 2011.05.05 20:37 | sell | 0.26 | eurusd | 1.45609 | 1.45609 | 1.45174 | 2011.05.05 20:38 | 1.45609 | 0.00 | 0.00 | 0.00 | 0.00 |
| 115373930 | 2011.05.05 20:38 | sell | 0.26 | eurusd | 1.45595 | 1.45570 | 1.45160 | 2011.05.05 21:04 | 1.45547 | 0.00 | 0.00 | 0.00 | 12.48 |
| 115377382 | 2011.05.05 20:58 | sell | 0.26 | usdchf | 0.86921 | 0.95941 | 0.86913 | 2011.05.26 15:34 | 0.86913 | 0.00 | 0.00 | -3.66 | 2.39 |
| 115378725 | 2011.05.05 21:04 | sell | 0.26 | eurusd | 1.45540 | 1.45458 | 1.45105 | 2011.05.05 21:11 | 1.45458 | 0.00 | 0.00 | 0.00 | 21.32 |
| 115380315 | 2011.05.05 21:11 | sell | 0.26 | eurusd | 1.45442 | 1.54458 | 1.45008 | 2011.05.05 22:11 | 1.45259 | 0.00 | 0.00 | 0.00 | 47.58 |
| 115471983 | 2011.05.06 10:22 | sell | 1.07 | eurusd | 1.45556 | 1.45513 | 1.45116 | 2011.05.06 10:25 | 1.45513 | 0.00 | 0.00 | 0.00 | 46.01 |
| 115472306 | 2011.05.06 10:25 | sell | 0.20 | audusd | 1.06902 | 1.06784 | 1.06473 | 2011.05.06 10:29 | 1.06784 | 0.00 | 0.00 | 0.00 | 23.60 |
| 115472481 | 2011.05.06 10:26 | sell | 0.27 | eurusd | 1.45506 | 1.45269 | 1.45075 | 2011.05.06 10:34 | 1.45243 | 0.00 | 0.00 | 0.00 | 71.01 |
| 115473255 | 2011.05.06 10:29 | sell | 0.26 | audusd | 1.06780 | 1.06671 | 1.06350 | 2011.05.06 10:35 | 1.06671 | 0.00 | 0.00 | 0.00 | 28.34 |
| 115474957 | 2011.05.06 10:34 | sell | 0.27 | eurusd | 1.45239 | 1.54250 | 1.44800 | 2011.05.06 14:43 | 1.45160 | 0.00 | 0.00 | 0.00 | 21.33 |
| 115475228 | 2011.05.06 10:35 | sell | 0.26 | audusd | 1.06653 | 1.15676 | 1.06226 | 2011.05.06 11:14 | 1.06458 | 0.00 | 0.00 | 0.00 | 50.70 |
| 115477413 | 2011.05.06 10:40 | buy | 0.25 | usdchf | 0.87117 | 0.78098 | 0.87548 | 2011.05.06 11:15 | 0.87174 | 0.00 | 0.00 | 0.00 | 16.35 |
| 115478345 | 2011.05.06 10:41 | sell | 1.00 | gbpusd | 1.63966 | 1.63962 | 0.00000 | 2011.05.06 10:42 | 1.63958 | 0.00 | 0.00 | 0.00 | 8.00 |
| 115514261 | 2011.05.06 13:45 | buy | 0.27 | eurusd | 1.45329 | 1.45331 | 1.45764 | 2011.05.06 14:00 | 1.45331 | 0.00 | 0.00 | 0.00 | 0.54 |
| 115514265 | 2011.05.06 13:45 | sell | 0.27 | audusd | 1.06762 | 1.15779 | 1.06329 | 2011.05.06 14:43 | 1.06669 | 0.00 | 0.00 | 0.00 | 25.11 |
| 115515504 | 2011.05.06 13:54 | sell | 0.27 | gbpusd | 1.64223 | 1.64196 | 1.63792 | 2011.05.06 14:28 | 1.64196 | 0.00 | 0.00 | 0.00 | 7.29 |
| 115515509 | 2011.05.06 13:54 | sell | 0.27 | gbpusd | 1.64223 | 1.64197 | 1.63792 | 2011.05.06 14:28 | 1.64197 | 0.00 | 0.00 | 0.00 | 7.02 |
| 115522751 | 2011.05.06 14:43 | sell | 0.31 | gbpusd | 1.64168 | 1.73188 | 1.63738 | 2011.05.06 15:14 | 1.64071 | 0.00 | 0.00 | 0.00 | 30.07 |
| 115522761 | 2011.05.06 14:43 | buy | 0.29 | audusd | 1.06663 | 1.06745 | 1.07096 | 2011.05.06 14:58 | 1.06745 | 0.00 | 0.00 | 0.00 | 23.78 |
| 115522778 | 2011.05.06 14:43 | sell | 0.27 | eurusd | 1.45128 | 1.45061 | 1.44693 | 2011.05.06 14:49 | 1.45045 | 0.00 | 0.00 | 0.00 | 22.41 |
| 115524157 | 2011.05.06 14:49 | sell | 0.27 | eurusd | 1.45033 | 1.45032 | 1.44593 | 2011.05.06 14:50 | 1.45032 | 0.00 | 0.00 | 0.00 | 0.27 |
| 115524531 | 2011.05.06 14:50 | sell | 0.27 | eurusd | 1.45026 | 1.45005 | 1.44586 | 2011.05.06 14:51 | 1.45005 | 0.00 | 0.00 | 0.00 | 5.67 |
| 115524837 | 2011.05.06 14:52 | sell | 0.27 | eurusd | 1.44994 | 1.44960 | 1.44560 | 2011.05.06 15:30 | 1.44914 | 0.00 | 0.00 | 0.00 | 21.60 |
| 115525806 | 2011.05.06 14:58 | sell | 0.26 | audusd | 1.06742 | 1.15763 | 1.06313 | 2011.05.12 04:30 | 1.06313 | 0.00 | 0.00 | -29.94 | 111.54 |
| 115525810 | 2011.05.06 14:58 | buy | 0.26 | audusd | 1.06763 | 0.97742 | 1.07192 | 2011.05.06 15:14 | 1.06799 | 0.00 | 0.00 | 0.00 | 9.36 |
| 115529358 | 2011.05.06 15:15 | sell | 0.28 | gbpusd | 1.64049 | 1.73068 | 1.63618 | 2011.05.06 15:27 | 1.63999 | 0.00 | 0.00 | 0.00 | 14.00 |
| 115531547 | 2011.05.06 15:27 | buy | 0.26 | eurusd | 1.45030 | 1.45066 | 1.45463 | 2011.05.06 15:28 | 1.45066 | 0.00 | 0.00 | 0.00 | 9.36 |
| 115531856 | 2011.05.06 15:28 | sell | 0.28 | eurusd | 1.45065 | 1.54082 | 1.44632 | 2011.05.06 15:30 | 1.44974 | 0.00 | 0.00 | 0.00 | 25.48 |
| 115532521 | 2011.05.06 15:30 | sell | 0.27 | usdchf | 0.87506 | 0.87502 | 0.87125 | 2011.05.06 15:32 | 0.87502 | 0.00 | 0.00 | 0.00 | 1.23 |
| 115532865 | 2011.05.06 15:30 | buy | 0.28 | eurusd | 1.44894 | 1.45202 | 1.45316 | 2011.05.06 15:32 | 1.45202 | 0.00 | 0.00 | 0.00 | 86.24 |
| 115534863 | 2011.05.06 15:32 | sell | 0.28 | eurusd | 1.45194 | 1.45186 | 1.44765 | 2011.05.06 15:34 | 1.45186 | 0.00 | 0.00 | 0.00 | 2.24 |
| 115534872 | 2011.05.06 15:32 | buy | 0.28 | eurusd | 1.45215 | 1.45246 | 1.45644 | 2011.05.06 15:33 | 1.45246 | 0.00 | 0.00 | 0.00 | 8.68 |
| 115536114 | 2011.05.06 15:34 | sell | 0.30 | audusd | 1.07376 | 1.07305 | 1.06945 | 2011.05.06 15:36 | 1.07305 | 0.00 | 0.00 | 0.00 | 21.30 |
| 115536122 | 2011.05.06 15:34 | buy | 0.28 | eurusd | 1.45184 | 1.45228 | 1.45616 | 2011.05.06 15:57 | 1.45255 | 0.00 | 0.00 | 0.00 | 19.88 |
| 115536584 | 2011.05.06 15:35 | buy | 0.26 | eurusd | 1.44967 | 1.44973 | 1.45402 | 2011.05.06 15:35 | 1.44973 | 0.00 | 0.00 | 0.00 | 1.56 |
| 115537197 | 2011.05.06 15:35 | sell | 0.26 | gbpusd | 1.64032 | 1.63776 | 1.63604 | 2011.05.06 15:38 | 1.63776 | 0.00 | 0.00 | 0.00 | 66.56 |
| 115538933 | 2011.05.06 15:38 | sell | 0.27 | audusd | 1.06953 | 1.06920 | 1.06527 | 2011.05.11 19:49 | 1.06881 | 0.00 | 0.00 | -15.48 | 19.44 |
| 115538950 | 2011.05.06 15:38 | sell | 0.26 | gbpusd | 1.63766 | 1.63763 | 1.63336 | 2011.05.06 15:40 | 1.63763 | 0.00 | 0.00 | 0.00 | 0.78 |
| 115539906 | 2011.05.06 15:40 | sell | 0.25 | gbpusd | 1.63760 | 1.72781 | 1.63331 | 2011.05.09 13:58 | 1.63638 | 0.00 | 0.00 | -0.90 | 30.50 |
| 115542852 | 2011.05.06 15:46 | buy | 0.23 | eurusd | 1.45041 | 1.45076 | 1.45473 | 2011.05.06 15:47 | 1.45076 | 0.00 | 0.00 | 0.00 | 8.05 |
| 115543339 | 2011.05.06 15:47 | sell | 0.23 | gbpusd | 1.64050 | 1.63975 | 1.63620 | 2011.05.06 15:50 | 1.63975 | 0.00 | 0.00 | 0.00 | 17.25 |
| 115544535 | 2011.05.06 15:50 | sell | 0.23 | gbpusd | 1.63967 | 1.72987 | 1.63537 | 2011.05.09 00:14 | 1.63729 | 0.00 | 0.00 | -0.83 | 54.74 |
| 115546727 | 2011.05.06 15:56 | buy | 0.21 | eurusd | 1.45128 | 1.45136 | 1.45563 | 2011.05.06 15:57 | 1.45136 | 0.00 | 0.00 | 0.00 | 1.68 |
| 115546914 | 2011.05.06 15:57 | sell | 0.21 | eurusd | 1.45163 | 1.54179 | 1.44729 | 2011.05.06 18:56 | 1.44729 | 0.00 | 0.00 | 0.00 | 91.14 |
| 115551365 | 2011.05.06 16:08 | sell | 0.21 | eurusd | 1.45384 | 1.54402 | 1.44952 | 2011.05.06 16:48 | 1.45212 | 0.00 | 0.00 | 0.00 | 36.12 |
| 115659032 | 2011.05.09 00:18 | sell | 0.96 | eurusd | 1.43717 | 1.52751 | 1.43301 | 2011.05.09 15:36 | 1.43630 | 0.00 | 0.00 | 0.00 | 83.52 |
| 115708556 | 2011.05.09 09:04 | buy | 1.00 | eurusd | 1.43994 | 0.00000 | 1.44013 | 2011.05.09 09:07 | 1.44013 | 0.00 | 0.00 | 0.00 | 19.00 |
| 115754950 | 2011.05.09 13:58 | buy | 0.21 | usdchf | 0.87690 | 0.87742 | 0.88125 | 2011.05.09 15:37 | 0.87772 | 0.00 | 0.00 | 0.00 | 19.62 |
| 115768770 | 2011.05.09 15:37 | buy | 0.30 | gbpusd | 1.63305 | 1.63321 | 1.63736 | 2011.05.09 15:41 | 1.63321 | 0.00 | 0.00 | 0.00 | 4.80 |
| 115769464 | 2011.05.09 15:41 | sell | 0.30 | gbpusd | 1.63319 | 1.72340 | 1.62890 | 2011.05.09 15:49 | 1.63248 | 0.00 | 0.00 | 0.00 | 21.30 |
| 115769465 | 2011.05.09 15:41 | buy | 0.30 | gbpusd | 1.63340 | 1.54319 | 1.63769 | 2011.05.09 21:10 | 1.63769 | 0.00 | 0.00 | 0.00 | 128.70 |
| 115771737 | 2011.05.09 15:50 | buy | 0.28 | gbpusd | 1.63219 | 1.63229 | 1.63650 | 2011.05.09 15:55 | 1.63229 | 0.00 | 0.00 | 0.00 | 2.80 |
| 115772298 | 2011.05.09 15:53 | buy | 0.25 | eurusd | 1.43209 | 1.43238 | 1.43640 | 2011.05.09 15:57 | 1.43238 | 0.00 | 0.00 | 0.00 | 7.25 |
| 115773288 | 2011.05.09 15:57 | buy | 0.28 | gbpusd | 1.63244 | 1.63250 | 1.63676 | 2011.05.09 15:58 | 1.63250 | 0.00 | 0.00 | 0.00 | 1.68 |
| 115773294 | 2011.05.09 15:57 | sell | 0.25 | eurusd | 1.43216 | 1.43214 | 1.42783 | 2011.05.09 16:10 | 1.43214 | 0.00 | 0.00 | 0.00 | 0.50 |
| 115775493 | 2011.05.09 16:10 | buy | 0.28 | usdchf | 0.87738 | 0.87740 | 0.88167 | 2011.05.09 16:15 | 0.87740 | 0.00 | 0.00 | 0.00 | 0.64 |
| 115775501 | 2011.05.09 16:10 | buy | 0.27 | eurusd | 1.43236 | 1.34221 | 1.43671 | 2011.05.09 23:07 | 1.43671 | 0.00 | 0.00 | 0.00 | 117.45 |
| 115776866 | 2011.05.09 16:15 | buy | 0.25 | gbpusd | 1.63169 | 1.63197 | 1.63598 | 2011.05.09 16:17 | 1.63223 | 0.00 | 0.00 | 0.00 | 13.50 |
| 116269564 | 2011.05.11 19:36 | sell | 0.10 | gbpusd | 1.63741 | 1.63673 | 1.63313 | 2011.05.11 19:46 | 1.63673 | 0.00 | 0.00 | 0.00 | 6.80 |
| 116269570 | 2011.05.11 19:36 | buy | 0.10 | gbpusd | 1.63763 | 1.63772 | 1.64191 | 2011.05.11 19:39 | 1.63772 | 0.00 | 0.00 | 0.00 | 0.90 |
| 116269595 | 2011.05.11 19:36 | buy | 1.01 | eurusd | 1.42193 | 1.42215 | 1.42627 | 2011.05.11 19:39 | 1.42215 | 0.00 | 0.00 | 0.00 | 22.22 |
| 116270182 | 2011.05.11 19:39 | sell | 0.30 | eurusd | 1.42209 | 1.42189 | 1.41774 | 2011.05.11 19:40 | 1.42189 | 0.00 | 0.00 | 0.00 | 6.00 |
| 116270468 | 2011.05.11 19:40 | buy | 0.30 | usdchf | 0.88613 | 0.88640 | 0.89043 | 2011.05.11 19:46 | 0.88640 | 0.00 | 0.00 | 0.00 | 9.14 |
| 116270473 | 2011.05.11 19:40 | buy | 0.30 | eurusd | 1.42190 | 1.42191 | 1.42628 | 2011.05.12 10:48 | 1.42191 | 0.00 | 0.00 | 0.54 | 0.30 |
| 116271849 | 2011.05.11 19:46 | buy | 0.29 | gbpusd | 1.63689 | 1.54668 | 1.63732 | 2011.05.26 16:39 | 1.63732 | 0.00 | 0.00 | -2.01 | 12.47 |
| 116293685 | 2011.05.11 22:08 | sell | 0.28 | audusd | 1.06932 | 1.15962 | 1.06512 | 2011.05.12 04:30 | 1.06512 | 0.00 | 0.00 | -16.21 | 117.60 |
| 116361871 | 2011.05.12 08:59 | sell | 0.30 | gbpusd | 1.63480 | 1.63429 | 1.63061 | 2011.05.12 09:09 | 1.63429 | 0.00 | 0.00 | 0.00 | 15.30 |
| 116361876 | 2011.05.12 08:59 | sell | 0.30 | audusd | 1.06162 | 1.06159 | 1.05729 | 2011.05.12 09:11 | 1.06159 | 0.00 | 0.00 | 0.00 | 0.90 |
| 116364175 | 2011.05.12 09:11 | sell | 0.30 | audusd | 1.06145 | 1.06104 | 1.05712 | 2011.05.12 09:16 | 1.06104 | 0.00 | 0.00 | 0.00 | 12.30 |
| 116364980 | 2011.05.12 09:16 | sell | 0.30 | audusd | 1.06086 | 1.06013 | 1.05656 | 2011.05.12 09:20 | 1.06013 | 0.00 | 0.00 | 0.00 | 21.90 |
| 116366135 | 2011.05.12 09:20 | sell | 0.30 | eurusd | 1.41908 | 1.41876 | 1.41470 | 2011.05.12 12:13 | 1.41876 | 0.00 | 0.00 | 0.00 | 9.60 |
| 116366147 | 2011.05.12 09:20 | sell | 0.30 | audusd | 1.05997 | 1.15016 | 1.05566 | 2011.05.12 12:33 | 1.05902 | 0.00 | 0.00 | 0.00 | 28.50 |
| 116381741 | 2011.05.12 10:42 | buy | 0.28 | usdchf | 0.88644 | 0.88703 | 0.89080 | 2011.05.12 10:45 | 0.88703 | 0.00 | 0.00 | 0.00 | 18.62 |
| 116382217 | 2011.05.12 10:45 | sell | 0.13 | usdcad | 0.96443 | 0.96427 | 0.96021 | 2011.05.12 10:52 | 0.96427 | 0.00 | 0.00 | 0.00 | 2.16 |
| 116383045 | 2011.05.12 10:52 | buy | 0.19 | eurusd | 1.42285 | 1.42292 | 1.42723 | 2011.05.12 11:02 | 1.42292 | 0.00 | 0.00 | 0.00 | 1.33 |
| 116383048 | 2011.05.12 10:52 | sell | 0.19 | usdcad | 0.96387 | 0.96382 | 0.95962 | 2011.05.12 11:02 | 0.96382 | 0.00 | 0.00 | 0.00 | 0.99 |
| 116384667 | 2011.05.12 11:02 | buy | 0.19 | usdchf | 0.88622 | 0.88630 | 0.89051 | 2011.05.12 11:04 | 0.88630 | 0.00 | 0.00 | 0.00 | 1.71 |
| 116384811 | 2011.05.12 11:03 | sell | 0.14 | nzdusd | 0.78952 | 0.78935 | 0.78532 | 2011.05.12 11:11 | 0.78905 | 0.00 | 0.00 | 0.00 | 6.58 |
| 116386002 | 2011.05.12 11:11 | sell | 0.11 | nzdusd | 0.78875 | 0.78870 | 0.78455 | 2011.05.12 12:09 | 0.78810 | 0.00 | 0.00 | 0.00 | 7.15 |
| 116400098 | 2011.05.12 12:10 | buy | 0.15 | nzdusd | 0.78812 | 0.69782 | 0.79232 | 2011.05.12 17:40 | 0.79232 | 0.00 | 0.00 | 0.00 | 63.00 |
| 116400891 | 2011.05.12 12:13 | sell | 0.16 | eurusd | 1.41862 | 1.50878 | 1.41428 | 2011.05.12 12:34 | 1.41680 | 0.00 | 0.00 | 0.00 | 29.12 |
| 116403495 | 2011.05.12 12:25 | sell | 0.22 | usdcad | 0.96516 | 1.05537 | 0.96087 | 2011.05.13 10:28 | 0.96087 | 0.00 | 0.00 | 0.25 | 98.22 |
| 116405938 | 2011.05.12 12:34 | buy | 0.16 | gbpusd | 1.62902 | 1.62960 | 1.63337 | 2011.05.12 12:36 | 1.62966 | 0.00 | 0.00 | 0.00 | 10.24 |
| 116405942 | 2011.05.12 12:34 | sell | 0.15 | eurusd | 1.41689 | 1.50699 | 1.41249 | 2011.05.12 13:08 | 1.41249 | 0.00 | 0.00 | 0.00 | 66.00 |
| 116405947 | 2011.05.12 12:34 | sell | 0.15 | audusd | 1.05882 | 1.14899 | 1.05449 | 2011.05.13 19:14 | 1.05449 | 0.00 | 0.00 | -2.82 | 64.95 |
| 116725865 | 2011.05.16 00:02 | buy | 0.16 | audusd | 1.05799 | 1.05819 | 1.06200 | 2011.05.16 10:12 | 1.05819 | 0.00 | 0.00 | 0.00 | 3.20 |
| 116725938 | 2011.05.16 00:03 | buy | 0.16 | nzdusd | 0.78549 | 0.69489 | 0.78939 | 2011.05.16 00:15 | 0.78659 | 0.00 | 0.00 | 0.00 | 17.60 |
| 116726725 | 2011.05.16 00:12 | buy | 0.15 | nzdusd | 0.78800 | 0.69630 | 0.79080 | 2011.05.18 06:28 | 0.79080 | 0.00 | 0.00 | 0.92 | 42.00 |
| 116726732 | 2011.05.16 00:12 | buy | 0.15 | eurusd | 1.40914 | 1.41116 | 1.41328 | 2011.05.16 09:55 | 1.41116 | 0.00 | 0.00 | 0.00 | 30.30 |
| 116780058 | 2011.05.16 09:55 | buy | 1.01 | eurusd | 1.41154 | 1.41166 | 1.41589 | 2011.05.16 09:59 | 1.41166 | 0.00 | 0.00 | 0.00 | 12.12 |
| 116780672 | 2011.05.16 09:59 | sell | 0.46 | eurusd | 1.41165 | 1.41163 | 1.40732 | 2011.05.16 12:10 | 1.41119 | 0.00 | 0.00 | 0.00 | 21.16 |
| 116782451 | 2011.05.16 10:10 | sell | 0.31 | gbpusd | 1.61941 | 1.70976 | 1.61526 | 2011.05.16 12:49 | 1.61837 | 0.00 | 0.00 | 0.00 | 32.24 |
| 116782495 | 2011.05.16 10:10 | sell | 0.85 | eurusd | 1.41203 | 1.50220 | 1.40770 | 2011.05.16 12:09 | 1.41184 | 0.00 | 0.00 | 0.00 | 16.15 |
| 116782815 | 2011.05.16 10:12 | sell | 0.11 | audusd | 1.05830 | 1.05773 | 1.05401 | 2011.05.16 11:18 | 1.05743 | 0.00 | 0.00 | 0.00 | 9.57 |
| 116782817 | 2011.05.16 10:12 | buy | 0.11 | audusd | 1.05851 | 1.05867 | 1.06280 | 2011.05.16 10:33 | 1.05888 | 0.00 | 0.00 | 0.00 | 4.07 |
| 116786050 | 2011.05.16 10:32 | buy | 0.31 | audusd | 1.05909 | 1.05960 | 1.06342 | 2011.05.16 16:41 | 1.05960 | 0.00 | 0.00 | 0.00 | 15.81 |
| 116793223 | 2011.05.16 11:18 | sell | 0.10 | audusd | 1.05725 | 1.14748 | 1.05298 | 2011.05.16 11:56 | 1.05637 | 0.00 | 0.00 | 0.00 | 8.80 |
| 116800200 | 2011.05.16 11:54 | buy | 0.27 | usdchf | 0.88820 | 0.79805 | 0.89255 | 2011.05.16 12:05 | 0.88885 | 0.00 | 0.00 | 0.00 | 19.74 |
| 116802238 | 2011.05.16 12:04 | buy | 0.27 | audusd | 1.05621 | 1.05644 | 1.06051 | 2011.05.16 16:08 | 1.05644 | 0.00 | 0.00 | 0.00 | 6.21 |
| 116803518 | 2011.05.16 12:09 | sell | 0.12 | nzdusd | 0.77980 | 0.87014 | 0.77564 | 2011.05.16 13:10 | 0.77839 | 0.00 | 0.00 | 0.00 | 16.92 |
| 116803591 | 2011.05.16 12:10 | buy | 0.13 | eurusd | 1.41120 | 1.41131 | 1.41554 | 2011.05.16 12:15 | 1.41131 | 0.00 | 0.00 | 0.00 | 1.43 |
| 116810586 | 2011.05.16 12:51 | sell | 0.14 | usdchf | 0.88829 | 0.97849 | 0.88399 | 2011.05.16 16:08 | 0.88567 | 0.00 | 0.00 | 0.00 | 41.41 |
| 116810953 | 2011.05.16 12:53 | sell | 0.13 | eurusd | 1.41087 | 1.41081 | 1.40652 | 2011.05.16 12:56 | 1.41081 | 0.00 | 0.00 | 0.00 | 0.78 |
| 116811476 | 2011.05.16 12:56 | buy | 0.13 | eurusd | 1.41090 | 1.41243 | 1.41523 | 2011.05.16 13:34 | 1.41264 | 0.00 | 0.00 | 0.00 | 22.62 |
| 116828482 | 2011.05.16 14:45 | sell | 0.13 | eurusd | 1.41098 | 1.50116 | 1.40666 | 2011.05.23 07:29 | 1.40666 | 0.00 | 0.00 | -5.97 | 56.16 |
| 116845317 | 2011.05.16 16:08 | buy | 0.14 | eurusd | 1.41676 | 1.41678 | 1.42109 | 2011.05.16 16:10 | 1.41678 | 0.00 | 0.00 | 0.00 | 0.28 |
| 116846724 | 2011.05.16 16:13 | buy | 0.27 | eurusd | 1.41666 | 1.41756 | 1.42100 | 2011.05.16 16:24 | 1.41756 | 0.00 | 0.00 | 0.00 | 24.30 |
| 116846897 | 2011.05.16 16:14 | sell | 0.16 | usdchf | 0.88556 | 0.88480 | 0.88129 | 2011.05.16 16:24 | 0.88480 | 0.00 | 0.00 | 0.00 | 13.74 |
| 116848972 | 2011.05.16 16:24 | buy | 0.17 | audusd | 1.05735 | 1.05958 | 1.06165 | 2011.05.16 16:41 | 1.05958 | 0.00 | 0.00 | 0.00 | 37.91 |
| 116853399 | 2011.05.16 16:40 | sell | 0.27 | eurusd | 1.41905 | 1.41876 | 1.41473 | 2011.05.16 16:46 | 1.41876 | 0.00 | 0.00 | 0.00 | 7.83 |
| 116853644 | 2011.05.16 16:41 | sell | 0.14 | nzdusd | 0.78024 | 0.78005 | 0.77608 | 2011.05.16 16:52 | 0.78005 | 0.00 | 0.00 | 0.00 | 2.66 |
| 116853726 | 2011.05.16 16:41 | sell | 0.15 | audusd | 1.05961 | 1.05944 | 1.05534 | 2011.05.16 16:50 | 1.05944 | 0.00 | 0.00 | 0.00 | 2.55 |
| 116855419 | 2011.05.16 16:50 | sell | 0.16 | audusd | 1.05927 | 1.05923 | 1.05495 | 2011.05.16 16:52 | 1.05923 | 0.00 | 0.00 | 0.00 | 0.64 |
| 116855848 | 2011.05.16 16:52 | sell | 0.16 | audusd | 1.05894 | 1.05881 | 1.05460 | 2011.05.16 22:48 | 1.05849 | 0.00 | 0.00 | 0.00 | 7.20 |
| 116863030 | 2011.05.16 17:18 | sell | 1.20 | eurusd | 1.42079 | 1.51095 | 1.41645 | 2011.05.16 21:18 | 1.42063 | 0.00 | 0.00 | 0.00 | 19.20 |
| 116863094 | 2011.05.16 17:18 | sell | 0.11 | gbpusd | 1.62206 | 1.62192 | 1.61776 | 2011.05.16 17:35 | 1.62192 | 0.00 | 0.00 | 0.00 | 1.54 |
| 116866423 | 2011.05.16 17:31 | sell | 0.11 | nzdusd | 0.78154 | 0.78118 | 0.77740 | 2011.05.17 10:25 | 0.78118 | 0.00 | 0.00 | -0.78 | 3.96 |
| 116869887 | 2011.05.16 17:50 | sell | 0.11 | gbpusd | 1.62161 | 1.62033 | 1.61728 | 2011.05.16 22:47 | 1.61984 | 0.00 | 0.00 | 0.00 | 19.47 |
| 116876529 | 2011.05.16 18:15 | sell | 0.27 | gbpusd | 1.62351 | 1.62334 | 1.61930 | 2011.05.16 19:10 | 1.62285 | 0.00 | 0.00 | 0.00 | 17.82 |
| 116876761 | 2011.05.16 18:17 | sell | 0.10 | audusd | 1.06282 | 1.06257 | 1.05852 | 2011.05.16 18:58 | 1.06219 | 0.00 | 0.00 | 0.00 | 6.30 |
| 116881634 | 2011.05.16 18:50 | buy | 0.10 | usdchf | 0.88217 | 0.88239 | 0.88652 | 2011.05.16 21:28 | 0.88286 | 0.00 | 0.00 | 0.00 | 7.82 |
| 116903305 | 2011.05.16 21:14 | buy | 0.28 | audusd | 1.06134 | 0.97104 | 1.06554 | 2011.05.18 06:26 | 1.06554 | 0.00 | 0.00 | 4.54 | 117.60 |
| 116903585 | 2011.05.16 21:17 | buy | 0.28 | gbpusd | 1.62367 | 1.62370 | 1.62783 | 2011.05.17 10:27 | 1.62370 | 0.00 | 0.00 | -0.11 | 0.84 |
| 116905319 | 2011.05.16 21:28 | buy | 0.42 | eurusd | 1.41927 | 1.32913 | 1.42363 | 2011.05.16 21:48 | 1.41942 | 0.00 | 0.00 | 0.00 | 6.30 |
| 116908233 | 2011.05.16 21:48 | sell | 1.09 | eurusd | 1.41952 | 1.50966 | 1.41516 | 2011.05.16 22:46 | 1.41763 | 0.00 | 0.00 | 0.00 | 206.01 |
| 116908235 | 2011.05.16 21:48 | buy | 1.09 | eurusd | 1.41966 | 1.32952 | 1.42402 | 2011.05.16 21:49 | 1.41985 | 0.00 | 0.00 | 0.00 | 20.71 |
| 116914391 | 2011.05.16 22:47 | sell | 0.14 | usdchf | 0.88390 | 0.97414 | 0.87964 | 2011.05.18 01:48 | 0.87964 | 0.00 | 0.00 | -0.19 | 67.80 |
| 116914992 | 2011.05.16 22:51 | sell | 0.28 | usdcad | 0.97398 | 1.06424 | 0.96974 | 2011.05.17 11:33 | 0.97350 | 0.00 | 0.00 | 0.32 | 13.81 |
| 116995553 | 2011.05.17 10:10 | buy | 0.17 | usdchf | 0.88628 | 0.88628 | 0.89055 | 2011.05.17 10:16 | 0.88628 | 0.00 | 0.00 | 0.00 | 0.00 |
| 117028442 | 2011.05.17 12:18 | buy | 0.29 | usdcad | 0.97331 | 0.88308 | 0.97758 | 2011.05.17 12:57 | 0.97385 | 0.00 | 0.00 | 0.00 | 16.08 |
| 117038870 | 2011.05.17 13:10 | buy | 0.29 | usdcad | 0.97350 | 0.88323 | 0.97773 | 2011.05.17 14:43 | 0.97377 | 0.00 | 0.00 | 0.00 | 8.04 |
| 117093983 | 2011.05.17 17:18 | sell | 0.29 | usdchf | 0.88597 | 0.97619 | 0.88169 | 2011.05.17 17:29 | 0.88563 | 0.00 | 0.00 | 0.00 | 11.13 |
| 117673306 | 2011.05.20 20:14 | buy | 1.52 | eurusd | 1.41896 | 1.32880 | 1.42330 | 2011.05.20 20:18 | 1.41964 | 0.00 | 0.00 | 0.00 | 103.36 |
| 117673316 | 2011.05.20 20:14 | buy | 0.14 | audusd | 1.06847 | 1.06916 | 1.07279 | 2011.05.20 20:20 | 1.06916 | 0.00 | 0.00 | 0.00 | 9.66 |
| 117673319 | 2011.05.20 20:14 | buy | 0.13 | nzdusd | 0.79775 | 0.79814 | 0.80193 | 2011.05.20 20:21 | 0.79842 | 0.00 | 0.00 | 0.00 | 8.71 |
| 117673783 | 2011.05.20 20:18 | buy | 0.29 | eurusd | 1.41967 | 1.42051 | 1.42400 | 2011.05.20 20:20 | 1.42051 | 0.00 | 0.00 | 0.00 | 24.36 |
| 117674082 | 2011.05.20 20:20 | sell | 0.13 | audusd | 1.06912 | 1.06843 | 1.06483 | 2011.05.20 22:18 | 1.06843 | 0.00 | 0.00 | 0.00 | 8.97 |
| 117674086 | 2011.05.20 20:20 | buy | 0.13 | audusd | 1.06933 | 1.07026 | 1.07362 | 2011.05.20 20:23 | 1.07055 | 0.00 | 0.00 | 0.00 | 15.86 |
| 117674241 | 2011.05.20 20:21 | buy | 0.13 | nzdusd | 0.79875 | 0.79905 | 0.80295 | 2011.05.20 20:25 | 0.79905 | 0.00 | 0.00 | 0.00 | 3.90 |
| 117675431 | 2011.05.20 20:25 | buy | 0.20 | nzdusd | 0.79939 | 0.70905 | 0.80355 | 2011.05.24 15:02 | 0.80052 | 0.00 | 0.00 | 1.22 | 22.60 |
| 117687225 | 2011.05.20 22:13 | sell | 0.18 | gbpusd | 1.62696 | 1.62619 | 1.62262 | 2011.05.20 22:39 | 1.62588 | 0.00 | 0.00 | 0.00 | 19.44 |
| 117688501 | 2011.05.20 22:30 | sell | 0.15 | nzdusd | 0.79737 | 0.88777 | 0.79327 | 2011.05.20 23:10 | 0.79650 | 0.00 | 0.00 | 0.00 | 13.05 |
| 117692861 | 2011.05.20 23:10 | buy | 0.18 | nzdusd | 0.79637 | 0.70597 | 0.80047 | 2011.05.24 13:56 | 0.79841 | 0.00 | 0.00 | 1.10 | 36.72 |
| 117693346 | 2011.05.20 23:15 | sell | 0.30 | eurusd | 1.41431 | 1.50448 | 1.40998 | 2011.05.23 02:15 | 1.40998 | 0.00 | 0.00 | -1.92 | 129.90 |
| 117972292 | 2011.05.24 13:57 | sell | 1.32 | eurusd | 1.41113 | 1.50129 | 1.40679 | 2011.05.24 14:05 | 1.40960 | 0.00 | 0.00 | 0.00 | 201.96 |
| 117974129 | 2011.05.24 14:05 | sell | 0.31 | audusd | 1.05540 | 1.05513 | 1.05107 | 2011.05.24 14:21 | 1.05484 | 0.00 | 0.00 | 0.00 | 17.36 |
| 117975732 | 2011.05.24 14:16 | sell | 0.31 | eurusd | 1.40868 | 1.40846 | 1.40430 | 2011.05.24 14:21 | 1.40817 | 0.00 | 0.00 | 0.00 | 15.81 |
| 117976503 | 2011.05.24 14:20 | sell | 0.55 | nzdusd | 0.79780 | 0.88814 | 0.79364 | 2011.05.24 16:49 | 0.79700 | 0.00 | 0.00 | 0.00 | 44.00 |
| 117976523 | 2011.05.24 14:20 | sell | 0.15 | usdchf | 0.88039 | 0.97059 | 0.87609 | 2011.05.24 16:21 | 0.87945 | 0.00 | 0.00 | 0.00 | 16.03 |
| 117985068 | 2011.05.24 15:22 | sell | 0.32 | audusd | 1.05741 | 1.14759 | 1.05309 | 2011.05.24 15:46 | 1.05680 | 0.00 | 0.00 | 0.00 | 19.52 |
| 117994915 | 2011.05.24 16:21 | buy | 0.17 | eurusd | 1.41143 | 1.41149 | 1.41578 | 2011.05.24 16:56 | 1.41149 | 0.00 | 0.00 | 0.00 | 1.02 |
| 117996243 | 2011.05.24 16:27 | buy | 0.32 | eurusd | 1.40964 | 1.40991 | 1.41404 | 2011.05.24 16:51 | 1.41012 | 0.00 | 0.00 | 0.00 | 15.36 |
| 117999324 | 2011.05.24 16:49 | buy | 0.37 | usdchf | 0.87974 | 0.87974 | 0.88404 | 2011.05.24 17:07 | 0.87974 | 0.00 | 0.00 | 0.00 | 0.00 |
| 117999344 | 2011.05.24 16:49 | buy | 0.16 | usdjpy | 82.117 | 82.119 | 82.553 | 2011.05.24 17:01 | 82.119 | 0.00 | 0.00 | 0.00 | 0.39 |
| 117999346 | 2011.05.24 16:49 | buy | 0.15 | euraud | 1.33545 | 1.33558 | 1.33945 | 2011.05.24 17:59 | 1.33558 | 0.00 | 0.00 | 0.00 | 2.06 |
| 117999355 | 2011.05.24 16:50 | buy | 0.15 | nzdusd | 0.79687 | 0.79691 | 0.80107 | 2011.05.24 16:58 | 0.79691 | 0.00 | 0.00 | 0.00 | 0.60 |
| 118000992 | 2011.05.24 16:58 | buy | 0.17 | nzdusd | 0.79717 | 0.79755 | 0.80135 | 2011.05.24 17:12 | 0.79755 | 0.00 | 0.00 | 0.00 | 6.46 |
| 118001769 | 2011.05.24 17:01 | sell | 0.17 | usdjpy | 82.113 | 82.109 | 81.681 | 2011.05.24 20:00 | 82.075 | 0.00 | 0.00 | 0.00 | 7.87 |
| 118004926 | 2011.05.24 17:12 | buy | 0.18 | nzdusd | 0.79787 | 0.79794 | 0.80203 | 2011.05.24 17:18 | 0.79794 | 0.00 | 0.00 | 0.00 | 1.26 |
| 118006061 | 2011.05.24 17:17 | buy | 0.32 | euraud | 1.33466 | 1.24417 | 1.33867 | 2011.05.24 17:21 | 1.33505 | 0.00 | 0.00 | 0.00 | 13.20 |
| 118006235 | 2011.05.24 17:18 | buy | 0.16 | nzdusd | 0.79814 | 0.70781 | 0.79830 | 2011.05.24 18:32 | 0.79830 | 0.00 | 0.00 | 0.00 | 2.56 |
| 118013176 | 2011.05.24 17:59 | sell | 0.18 | euraud | 1.33557 | 1.42607 | 1.33464 | 2011.05.25 03:31 | 1.33464 | 0.00 | 0.00 | 0.86 | 17.66 |
| 118013178 | 2011.05.24 17:59 | buy | 0.18 | euraud | 1.33607 | 1.33656 | 1.34007 | 2011.05.24 18:12 | 1.33703 | 0.00 | 0.00 | 0.00 | 18.25 |
| 118017662 | 2011.05.24 18:19 | buy | 0.32 | euraud | 1.33748 | 1.24708 | 1.33856 | 2011.05.25 06:49 | 1.33856 | 0.00 | 0.00 | -7.46 | 36.27 |
| 118020104 | 2011.05.24 18:33 | buy | 0.17 | nzdusd | 0.79862 | 0.70829 | 0.80279 | 2011.05.26 04:11 | 0.80279 | 0.00 | 0.00 | 2.11 | 70.89 |
| 118032521 | 2011.05.24 19:59 | sell | 0.36 | euraud | 1.33594 | 1.42636 | 1.33186 | 2011.05.24 23:55 | 1.33553 | 0.00 | 0.00 | 0.00 | 15.59 |
| 118033081 | 2011.05.24 20:00 | sell | 0.18 | usdjpy | 82.057 | 82.047 | 81.627 | 2011.05.24 20:04 | 81.999 | 0.00 | 0.00 | 0.00 | 12.73 |
| 118034088 | 2011.05.24 20:08 | sell | 0.18 | usdjpy | 81.957 | 81.902 | 81.521 | 2011.05.24 20:33 | 81.857 | 0.00 | 0.00 | 0.00 | 21.99 |

| Ticket | Open Time | Type | Size | Item | Price | S/L | T/P | Close Time | Price | Commission | Taxes | Swap | Profit |
|---|---|---|---|---|---|---|---|---|---|---|---|---|---|
| 118037191 | 2011.05.24 20:33 | sell | 0.18 | usdjpy | 81.851 | 90.867 | 81.417 | 2011.05.26 15:45 | 81.417 | 0.00 | 0.00 | -0.44 | 95.95 |
| 118049208 | 2011.05.24 22:32 | buy | 0.32 | audusd | 1.05683 | 0.96653 | 1.06103 | 2011.05.26 08:59 | 1.06103 | 0.00 | 0.00 | 10.24 | 134.40 |
| 118091672 | 2011.05.25 08:25 | sell | 0.42 | euraud | 1.33949 | 1.42993 | 1.33543 | 2011.05.26 17:03 | 1.33543 | 0.00 | 0.00 | 5.95 | 180.63 |

Closed P/L: 5 683.62

Open Trades:

| Ticket | Open Time | Type | Size | Item | Price | S/L | T/P | Price | Commission | Taxes | Swap | Profit |
|---|---|---|---|---|---|---|---|---|---|---|---|---|
| 117677989 | 2011.05.20 20:45 | buy | 0.20 | audusd | 1.07058 | 0.98038 | 1.07488 | 1.06217 | 0.00 | 0.00 | 9.62 | -168.20 |
| 117678501 | 2011.05.20 20:48 | buy | 0.19 | eurusd | 1.42211 | 1.33199 | 1.42100 | 1.41300 | 0.00 | 0.00 | 1.04 | -173.09 |
| 118115987 | 2011.05.25 10:38 | sell | 0.28 | eurusd | 1.40284 | 1.49299 | 1.39849 | 1.41310 | 0.00 | 0.00 | -5.71 | -287.28 |
| 116384672 | 2011.05.12 11:02 | sell | 0.19 | usdcad | 0.96360 | 1.05381 | 0.96288 | 0.97876 | 0.00 | 0.00 | 3.14 | -294.29 |
| 116400092 | 2011.05.12 12:09 | sell | 0.15 | usdcad | 0.96452 | 1.05468 | 0.96018 | 0.97876 | 0.00 | 0.00 | 2.46 | -218.24 |
| 118432512 | 2011.05.26 21:17 | sell | 2.15 | euraud | 1.33025 | 1.42039 | 1.32589 | 1.33034 | 0.00 | 0.00 | 0.00 | -20.56 |
| 118432653 | 2011.05.26 21:18 | buy | 1.39 | usdjpy | 81.275 | 72.265 | 81.715 | 81.269 | 0.00 | 0.00 | 0.00 | -10.26 |
|  |  |  |  |  |  |  |  |  | 0.00 | 0.00 | 10.55 | -1 171.92 |

Floating P/L: -1 161.37

**Working Orders:**

| Ticket | Open Time | Type | Size | Item | Price | S / L | T / P | Market Price |
|--------|-----------|------|------|------|-------|-------|-------|--------------|
| | | | | | No transactions | | | |

**Summary:**

| Deposit/Withdrawal: | 5 000.00 | Credit Facility: | 0.00 | | |
|---|---|---|---|---|---|
| Closed Trade P/L: | 5 683.62 | Floating P/L: | -1 161.37 | Margin: | 5 377.11 |
| Balance: | 10 683.62 | Equity: | 9 522.25 | Free Margin: | 4 145.14 |

**Details:**

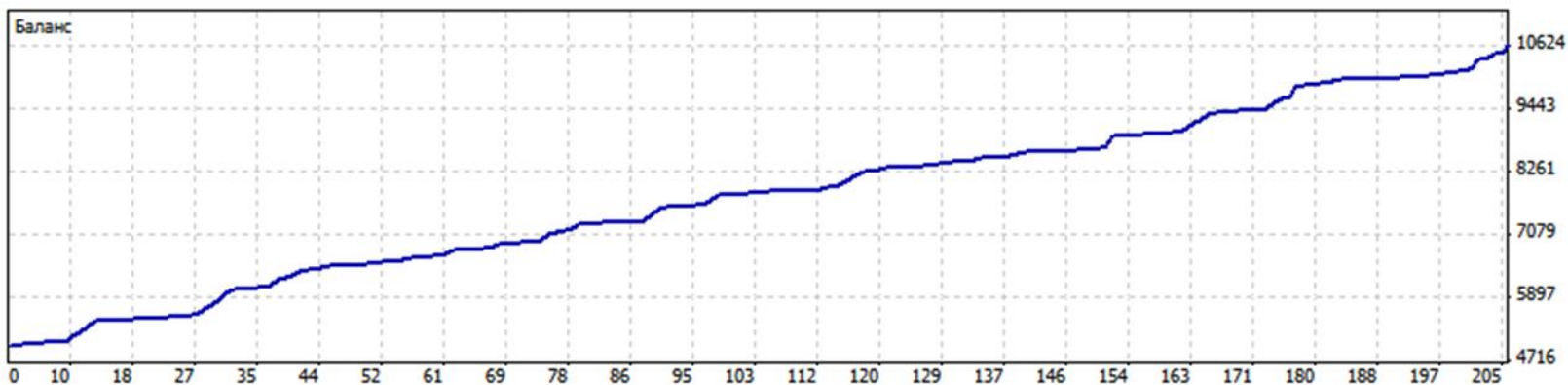

| Gross Profit: | 5 684.89 | Gross Loss: | 1.27 | Total Net Profit: | 5 683.62 |
|---|---|---|---|---|---|
| Profit Factor: | 4476.29 | Expected Payoff: | 27.72 | | |
| Absolute Drawdown: | 0.00 | Maximal Drawdown: | 1.27 (0.01%) | Relative Drawdown: | 0.01% (1.27) |
| Total Trades: | 205 | Short Positions (won %): | 122 (99.18%) | Long Positions (won %): | 83 (100.00%) |
| | | Profit Trades (% of total): | 204 (99.51%) | Loss trades (% of total): | 1 (0.49%) |
| Largest | | profit trade: | 206.01 | loss trade: | -1.27 |
| Average | profit trade: | 27.87 | | loss trade: | -1.27 |
| Maximum | consecutive wins ($): | 201 (5 392.34) | | consecutive losses ($): | 1 (-1.27) |
| Maximal | consecutive profit (count): | 5 392.34 (201) | | consecutive loss (count): | -1.27 (1) |
| Average | consecutive wins: | 102 | | consecutive losses: | 1 |